\documentclass[aps,superscriptaddress,twocolumn,prd]{revtex4-2}
\usepackage[T1]{fontenc}
\usepackage[utf8]{inputenc}
\usepackage{mathtools}
\usepackage{graphicx}
\usepackage{soul}
\usepackage{amsmath,amssymb}
\usepackage[colorlinks=true, pdfstartview=FitV, linkcolor=blue, citecolor=blue, urlcolor=magenta, breaklinks=true]{hyperref}

\usepackage{ae,aecompl}
\usepackage{hyperref}
\usepackage{amsmath}
\usepackage{amssymb}
\usepackage{bm}
\usepackage{cleveref}
\usepackage{multirow}

\usepackage{amssymb,amsmath,multirow,dcolumn,bm,float,graphicx,epstopdf}
\usepackage{calrsfs}  
\usepackage{epsfig}
\usepackage{soul}
\makeatletter

\usepackage{amssymb}

\usepackage{graphicx,color}
\usepackage[usenames,dvipsnames]{xcolor}

\usepackage{subfigure}

\usepackage{booktabs}
\usepackage{slashed}
\colorlet{darkgreen}{green!50!black}
\colorlet{brightyellow}{yellow!75!red}%
\colorlet{orange}{red!50!yellow}
\colorlet{darkblue}{blue!60!black}
\colorlet{darkred}{red!80!black}

\usepackage{soul}

\usepackage{mathtools}
\usepackage{mathrsfs}
\usepackage{bm}
\usepackage{relsize}
\usepackage{indentfirst}

\def\be{\begin{eqnarray} &&}

\def\ee{\end{eqnarray}}

\usepackage{hyperref}
\hypersetup{
    unicode=false,          
    pdftoolbar=true,        
    pdfmenubar=true,        
    pdffitwindow=false,     
    pdfstartview={FitH},    
    pdfnewwindow=true,      
    colorlinks=true,       
    linkcolor=red,          
    citecolor=red,        
    filecolor=cyan,         
    urlcolor=magenta        
}

\newcommand{\nc}{\newcommand}
\nc{\lamnr}{\lambda_{nr}}
\nc{\lamr}{\lambda_{r}}
\nc{\lamk}{\lambda_{k}}
\nc{\gp}{g({\bp})}
\nc{\gpp}{g({\bp'})}
\nc{\gpz}{g({\bp''})}
\nc{\gps}{g^{*}({\bp})}
\nc{\gpzs}{g^{*}({\bp''})}
\nc{\gpps}{g^{*}({\bp'})}
\nc{\bxi}{{\bf \xi}}
\nc{\bp}{{\bf p}}
\nc{\bpp}{{\bf p'}}
\nc{\bpz}{{\bf p''}}
\nc{\bk}{{\bf k}}
\nc{\bkp}{{\bf k'}}
\nc{\bkz}{{\bf k''}}
\nc{\bPi}{{\bf \Pi}}
\nc{\bera}{\langle}
\nc{\ket}{\rangle}
\nc{\bq}{{\bf q}}
\nc{\bqp}{{\bf q'}}
\nc{\tpi}{\tilde{\pi}}
\nc{\bpi}{\boldsymbol \pi}
\nc{\btpi}{\tilde{\boldsymbol \pi}}
\nc{\andre}[1]{\textcolor{red}{#1}}

\begin{document}
\title{Gluonic contributions to the pion parton distribution functions}

\author{Jiangshan Lan}
\email{jiangshanlan@impcas.ac.cn}
\affiliation{Institute of Modern Physics, Chinese Academy of Sciences, Lanzhou 730000, China}
\affiliation{School of Nuclear Physics, University of Chinese Academy of Sciences, Beijing, 100049, China}
\affiliation{CAS Key Laboratory of High Precision Nuclear Spectroscopy, Institute of Modern Physics, Chinese Academy of Sciences, Lanzhou 730000, China}
\affiliation{Advanced Energy Science and Technology Guangdong Laboratory, Huizhou, Guangdong 516000, China}

\author{Chandan Mondal}
\email{mondal@impcas.ac.cn}
\affiliation{Institute of Modern Physics, Chinese Academy of Sciences, Lanzhou 730000, China}
\affiliation{School of Nuclear Physics, University of Chinese Academy of Sciences, Beijing, 100049, China}
\affiliation{CAS Key Laboratory of High Precision Nuclear Spectroscopy, Institute of Modern Physics, Chinese Academy of Sciences, Lanzhou 730000, China}

\author{Xingbo Zhao}
\email{xbzhao@impcas.ac.cn}
\affiliation{Institute of Modern Physics, Chinese Academy of Sciences, Lanzhou 730000, China}
\affiliation{School of Nuclear Physics, University of Chinese Academy of Sciences, Beijing, 100049, China}
\affiliation{CAS Key Laboratory of High Precision Nuclear Spectroscopy, Institute of Modern Physics, Chinese Academy of Sciences, Lanzhou 730000, China}
\affiliation{Advanced Energy Science and Technology Guangdong Laboratory, Huizhou, Guangdong 516000, China}

\author{Tobias Frederico}

\email{tobias@ita.br}
\affiliation{Instituto Tecnol\'ogico de Aeron\'autica,  DCTA, 12228-900 S\~ao Jos\'e dos Campos,~Brazil}

\author{James P. Vary}
\email{jvary@iastate.edu}
\affiliation{Department of Physics and Astronomy, Iowa State University, Ames, IA 50011, USA}

\date{\today}

\begin{abstract}
We investigate the role of a dynamical gluon in the pion within the Basis Light-Front Quantization (BLFQ) framework and compare it with the solution of the Minkowski space Bethe-Salpeter equation (BSE), focusing on contributions beyond the valence sector.  We develop a framework to compute the quark-antiquark-gluon component of the pion, starting from its valence light-front wave function. Additionally, the quark and gluon parton distribution functions (PDFs) are derived by considering this higher Fock component in the pion state. The proposed operator, which acts on the valence state to produce the quark-antiquark-gluon component of the pion, is tested against results from BLFQ for its contribution to the quark and gluon PDFs, as well as with results from the BSE. In the BLFQ framework, we identify the effect of dynamical chiral symmetry breaking on the pion structure through the enhancement of the spin-flip matrix element. This enhancement is captured in the $|q\bar{q}g\rangle$ component of the light-front wave function and the associated gluon PDF. We explicitly demonstrate an enhancement of the low-$x$ contribution in the quark PDF, which is associated with the large spin-flip matrix element, which also drives the $\pi-\rho$ mass splitting. The proposed framework also enables an exploration of the effect of the dressed gluon mass for a given valence state at the pion scale. We show that when the gluon mass vanishes, departing from the constituent gluon picture, it has a significant impact on the gluon PDF.

\end{abstract}

\maketitle

\section{Introduction}

The light-front representation of the hadron~\cite{Brodsky:1997de,BakkerNPB2014} carries the full complexity of Quantum Chromodynamics (QCD), with their constituents, namely dressed quarks and gluons strongly interacting to build  an eigenstate of a mass squared operator. Such a description implies the dynamical coupling of an infinite set of  Fock components forming the hadron eigenstate. In practice a truncated Light-front Fock-space is adopted as a hadron  within Basis Light-Front Quantization (BLFQ)~\cite{Vary:2009gt}. To date, the valence  and valence plus one gluon  states were coupled to  describe, e.g. light-mesons~\cite{Lan:2021wok},  where the  confinement is introduced in the squared mass operator acting on  the valence sector and  the $|q\bar q g\rangle$ originates  from the  coupling with the $q\bar q$ sector through the  off-diagonal matrix elements of  the QCD LC Hamiltonian~\cite{Brodsky:1997de}, in such a way that the gluon plays an important dynamical role. Indeed, in the pion case the $q\bar q$  and $q\bar q g$ sectors each carry about  50\% 
 of  the total normalization~\cite{Lan:2021wok}, indicating the important role of the dynamical dressed gluon,  even in the presence of a confining interaction acting in the valence channel. We note that
 dynamical symmetry breaking should involve the coupling with an infinite number of Fock-states to dress the constituent quarks and at the same time provide the large splitting between the nearly massless pion (the Goldstone boson) and the rho meson. In this connection, the coupling of the valence state with the higher Fock-components can be cast into an effective interaction, as indicated by the  ``Iterative Resolvent method"~\cite{Brodsky:1997de}.  This effective interaction has been examined in Ref.~\cite{Burkardt98}, where it was proposed to enhance the spin flip matrix element of the effective quark-gluon coupling QCD LF-Hamiltonian by introducing a large  effective quark vertex  mass ($m_f$). This mechanism has been implemented with success in BLFQ to split the pion and rho meson masses~\cite{Lan:2021wok}.

 We should point out that the relevance of the higher Fock components of the pion state has been recently addressed phenomenologicaly with a new parametrization containing  $q\bar q$, $q\bar qq\bar q$, $q\bar q g$ and $q\bar q gg$ components in Ref.~\cite{Pasquini:2023aaf}, where those components  were fitted  simultaneously to describe the experimental data on the
pion PDFs~\cite{Barry:2018ort,NovikovPRD2020} and electromagnetic form factor.

 On the other hand, four-dimensional field theoretical approaches within the Dyson-Schwinger (DS) and Bethe-Salpeter (BS) frameworks describe  the pion as the Goldstone boson originated by the spontaneous breaking of the chiral symmetry in the light-quark sector.  These non-perturbative frameworks dress the light-quarks in the SU(3) flavor sector and split the pion and the rho meson as well as the kaon and $K^*$ meson (see e.g.~\cite{CloPPNP14,Eichmann:2016yit}). Those approaches have been formulated in Euclidean space, and the connection to the LF Fock expansion for any meson is not direct and demands some elaboration to access the parton distribution (see e.g.~\cite{Mezrag:2023nkp,Leitao:2017esb}, as well as in Lattice QCD~\cite{Roberts:2021nhw}). 
 
 However, the solution of the pion Bethe-Salpeter equation (BSE) in Minkowski space, like the one developed in Ref.~\cite{dePaula:2020qna} with massive constituent quarks and gluons, allows access to the valence component as well the inclusive contribution of the higher-Fock components to structure observables~\cite{dePaula:2022PRDL}. It was found that the pion admits a significant contribution from higher LF Fock-components with the valence carrying 70\% of the normalization.  The BSE in ladder approximation allows the coupling of the valence state with an infinite set
of Fock-components (see e.g.~\cite{Sales:1999ec,Cooke:1999yi,Cooke:2000ef,Sales:2001gk,Frederico:2010zh}). So far, one can only separate out the valence component of the wave function from the BS amplitude, by eliminating the relative LF time through the integration on the  LF energy, leaving only the longitudinal momentum fraction and transverse momentum, which  characterize the arguments  of valence wave function (see e.g. the detailed discussion in Refs.~\cite{dePaula:2020qna,MezragFBS}).

Ideally, BLFQ  and DS/BS frameworks applied to QCD should lead to the same results for the physical observables of a  given hadron.
However, truncations and insertions of confinement in different ways in the these approaches will, on their own, lead to different answers. Furthermore, in BLFQ the enhancement of the spin-flip matrix element is expected to have a distinctive hallmark on the pion structure at low-$x$ from the contribution of the $|q\bar q g\rangle$ Fock-component. 
We note that the BLFQ eigenvalue equations, when formulated without confinement and truncated to the $q\bar{q}g$ state, could eventually be related to the implementation of the Leibbrandt-Mandelstam gauge prescription in the ladder Bethe-Salpeter equation (BSE) when self-energy contributions are not considered~\cite{Liu:1992dg}. The development performed in that work, in principle, corresponds to the lowest-order kernel in the expansion of the light-front projected BSE (see, e.g., ~\cite{Sales:1999ec,Sales:2001gk,Frederico:2010zh}).

Our aim in this  paper is twofold: first, explore both the quark and gluon longitudinal momentum  fraction ($x$) distribution, or the PDFs, from the $q\bar qg$ component computed within a continuous approach, analyzing different parametrizations, without resorting to the discretization  adopted in the BLFQ method to describe the  pion; and, second, compare the BLFQ results with the BSE results for the  contribution to the quark PDF from the higher Fock sectors after subtraction of the  valence part from the total PDF.  Within this aim, we will determine the main characteristics of these contributions, particularly concerning the breaking of the symmetry around $x=1/2$. We note that the BSE model reproduces the pion experimental space-like electromagnetic form factor, as shown in Ref.~\cite{Ydrefors:2021dwa}. Furthermore, we will provide a practical method to generate the $q\bar{q}g$ contribution to the pion from the leading spin-antialigned valence wave function that can be used in a range of applications.

\section{Theoretical framework \label{Sec:PLFH}}

The bound state in LF field theory can be obtained by solving an eigenvalue problem of the Hamiltonian
in a frame with a vanishing total transverse momentum $(\vec P_\perp=0)$:
\begin{equation}
 P^-P^+|{\Psi}\rangle=M^2|{\Psi}\rangle\, ,
 \label{Eq:M2operator}
 \end{equation}
 where  $P^\pm=P^0 \pm P^3$ represent the LF Hamiltonian, $P^-$, and the longitudinal momentum, $P^+$, of the system, respectively. The eigenvalue $M^2$ is the mass squared of the bound state. 

The LF Hamiltonian we use contains the LF QCD Hamiltonian and  confinement, $P^-= P^-_{\rm QCD} +P^-_{\rm{C}}$~\cite{Lan:2021wok}. With one dynamical gluon, the LF QCD Hamiltonian in the LF gauge $A^+=0$~\cite{Brodsky:1997de,Lan:2021wok} reads
\begin{equation}
\label{Eq:PQCD}
\begin{aligned}
    P_{\rm QCD}^-=& \int \mathrm{d}x^-\mathrm{d}^2 x^{\perp} \Bigg[\frac{1}{2}\bar{\psi}\gamma^+\frac{m_{0}^2+(i\partial^\perp)^2}{i\partial^+}\psi\\
    & +\frac{1}{2}A_a^i\left[m_g^2+(i\partial^\perp)^2\right] A^i_a +g_s\bar{\psi}\gamma_{\mu}T^aA_a^{\mu}\psi \\
    &+\frac{1}{2}g_s^2\bar{\psi}\gamma^+T^a\psi\frac{1}{(i\partial^+)^2}\bar{\psi}\gamma^+T^a\psi \Bigg]\,,
\end{aligned}
\end{equation}
where $\psi$ and $A^\mu$ are the quark and gluon fields, respectively. $T^a$ is the half Gell-Mann matrix, $T^a=\lambda^a/2$, and $\gamma^+=\gamma^0+\gamma^3$, where $\gamma^\mu$ represents the Dirac matrix. The first two terms in Eq.~(\ref{Eq:PQCD}) are the kinetic energies of quark and gluon, while the last two terms describe their interactions with coupling constant $g_s$. $m_{0}$ and $m_g$ are the bare mass of quarks and the model gluon mass, respectively.  We acknowledge that the explicit gluon mass term $m_g^2$ in the LF Hamiltonian breaks gauge invariance, which is already compromised due to Fock space truncation. 
{In our framework, $m_g^2$ is introduced as an effective parameter to regularize infrared singularities and to model aspects of confinement. We note that in the Schwinger mechanism, the nonperturbative resummation of gluon self-energy corrections leads to a gauge-invariant, dynamically generated dressed gluon whose propagator behaves as if the gluon has an effective mass, even though its current (bare) mass remains zero~\cite{Cornwall:1981zr,Aguilar:2007nf,Binosi:2014aea}. Our treatment uses $m_g^2$ as a phenomenological input associated with a dressed gluon degree of freedom, consistent with similar infrared regularization strategies employed in Hamiltonian approaches and effective models.
}

Using the Fock sector dependent renormalization scheme~\cite{Karmanov:2008br,Li:2015iaw}, we introduce a mass counter term, $\delta m_{q}= m_0 -m_{q}$, in the leading Fock sector to regularize the quark self-energy. Here, $m_{q}$ is the renormalized quark mass. Apart from this, we introduce a different quark mass $m_f$ to parameterize the nonperturbative effects
in the vertex interactions~\cite{Burkardt98,Glazek:1992aq}. 
The confinement in the leading Fock sector includes transverse and longitudinal confining potentials~\cite{Lan:2021wok},
\begin{equation}
  \begin{split}\label{eqn:Hc}
  &P_{\rm C}^-P^+=\kappa^4\left\{x(1-x) \vec{r}_{\perp}^{\,2}-\frac{\partial_{x}[x(1-x)\partial_{x}]}{(m_q+m_{\bar{q}})^2}\right\}\,,
  \end{split}
\end{equation}
where $\kappa$ is the strength of the confinement, and $\vec r_{\perp}=\sqrt{x(1-x)}(\vec r_{\perp q}-\vec r_{ \perp\bar{q}})$ represents the holographic variable~\cite{Brodsky:2014yha}.

 In BLFQ, the longitudinal dynamics are described using a discretized plane-wave basis with momenta $p^+ = \frac{2\pi k}{L}$, where $k$ is half-integer (integer) for fermions (bosons), and the transverse dynamics use 2D harmonic oscillator (HO) wave functions $\phi_{n,m}(k_\perp; b)$ with quantum numbers $(n, m)$. The longitudinal motion is confined to a 1D box of length $2L$ with antiperiodic (periodic) boundary conditions for fermions (bosons), and the total longitudinal momentum is $P^+ = \frac{2\pi K}{L}$ with $K = \sum_i k_i$. The many-body basis is built from single-particle states $\{x, n, m, \lambda\}$, where $\lambda$ is helicity, and the total angular momentum projection is $m_J = \sum_i (m_i + \lambda_i)$. 

The basis truncations are controlled by $K$ (longitudinal) and $N_{\text{max}}$ (transverse), where $N_{\text{max}} \geq 2n_i + |m_i| + 1$ introduces UV and IR cutoffs, $\Lambda_{\text{UV}} = b\sqrt{N_{\text{max}}}$ and $\Lambda_{\text{IR}} = \frac{b}{\sqrt{N_{\text{max}}}}$. Numerical calculations use $N_{\text{max}} = 14$, $K = 15$, and the HO scale $b = 0.29$ GeV. Model parameters $\{m_q,\, m_g,\, \kappa,\, m_f,\, g_s  \}=\{0.39,\, 0.60,\, 0.65,\, 5.69,\, 1.92\}$ (GeV except $g_s$) are fitted to the mass spectra of unflavored light mesons~\cite{Lan:2021wok}.

The pion state vector obeys the eigenvalue equation~\eqref{Eq:M2operator} with $M^2\equiv M^2_\pi$, and
the state vector can be expressed on the null-plane, i.e.~$x^+=0$, through a Fock expansion of the form (see e.g.~\cite{Brodsky:1997de,CK_rev})
\begin{equation}
\label{Eq:Fock_expansion}
  \begin{aligned}
    |&\Psi\rangle = 
    \sum_{i_q s_q}\sum_{i_{\bar{q}} s_{\bar{q}}}\int \Bigl[\prod_{j=q, \bar{q}}\frac{d^3 p_j}{(2\pi)^32p^+_j}\Bigr]2P^+(2\pi)^3\\
    &\times \delta^{(3)}(\vec{p}_q + \vec{p}_{\bar{q}} - \vec{P})\psi^{(s_q, s_{\bar{q}})}_{q\bar{q};(i_q,i_{\bar{q}})}(x_q, \vec{p}_{q\perp}, x_{\bar{q}}, \vec{p}_{\bar{q}\perp}) \\
    & \times b^\dagger_{i_q s_q}(\vec{p}_q)d^\dagger_{i_{\bar{q}} s_{\bar{q}}}(\vec{p}_{\bar{q}})|0\rangle  + \sum_{i_q s_q}\sum_{i_{\bar{q}} s_{\bar{q}}}\sum_{\lambda a}\int\Bigl[\prod_{j=q, \bar{q}, g}\frac{d^3 p_j}{(2\pi)^32p^+_j}\Bigr] \\
    & \times 2P^+(2\pi)^3\delta^{(3)}(\vec{p}_q + \vec{p}_{\bar{q}} + \vec{p}_g - \vec{P})\psi^{(s_q,s_{\bar{q}},\lambda)}_{q\bar{q}g;(i_q,i_{\bar{q}},a)}(\lbrace x, \vec{p}_\perp \rbrace)\\
    & \times  b^\dagger_{i_q s_q}(\vec{p}_q)d^\dagger_{i_{\bar{q}} s_{\bar{q}}}(\vec{p}_{\bar{q}})a^\dagger_{\lambda a}(\vec{p}_g)|0\rangle + \cdots\,,    
   \end{aligned}
\end{equation}
where $\lbrace x, \vec{p}_\perp \rbrace \equiv \lbrace x_q, \vec{p}_{q\perp}, x_{\bar{q}}, \vec{p}_{\bar{q}\perp}, x_g, \vec{p}_{g\perp}  \rbrace$ satisfy the momentum conservation:
\begin{equation}\label{Eq:momcons}
\vec{p}_{q\perp}+\vec{p}_{\bar{q}\perp}+ \vec{p}_{g\perp}=0~~  \text{and}~~ x_q+ x_{\bar{q}}+ x_g=1  \,.
\end{equation}
Here $b^\dagger_{i_q s_q}$ $(d^\dagger_{i_{\bar{q}} s_{\bar{q}}})$ is the constituent (anti)quark creation operator,  $a^\dagger_{\lambda a}$ is the creation operator for the constituent gluon and  $i_q(i_{\bar q})$ and $a$ are color indices, $s_q$, $s_{\bar{q}}$, and $\lambda=\pm 1$ denote the helicity of the quark, antiquark and gluon, respectively.

The BLFQ  LF Hamiltonian contains confinement only in the leading Fock-sector, while the Hamiltonian in higher Fock-sector sectors is built from Eq.~\eqref{Eq:M2operator}
considering that the effective degrees of freedom are dressed quarks and gluons with constituent masses. Therefore, by inserting the Fock expansion \eqref{Eq:Fock_expansion} truncated at second order in ~\eqref{Eq:M2operator}, 
 one can derive the following equation for the  LF wave function of the $q\bar{q}g$ sector, which is also valid within the BLFQ approach:
\begin{equation}
\label{Eq:psi_qqg}
\psi^{(s_q,s_{\bar{q}},\lambda)}_{q\bar{q}g;(i_q,i_{\bar{q}},a)}
= 
   \frac{1}{M^2_\pi - M^2_{0,q\bar{q}g}}\Big[V\psi^{(s_q, s_{\bar{q}})}_{q\bar{q};(i_q,i_{\bar{q}})}
 \Big]\,,  
\end{equation}
where the mass-squared operator of the free $q\bar{q}g$ system reads
\begin{equation}
\label{Eq:M0_qqg}
  M^2_{0,q\bar{q}g} = \sum_{j=q, \bar{q}, g} \frac{\vec{p}^{\,2}_{j\perp} + m^2_j}{x_j}\,,
\end{equation}
and $V$ denotes the interaction connecting the $q\bar{q}$ and $q\bar{q}g$ sectors. In the present work, we truncate the Fock-space up to the $q\bar q g$ sector,  and the interaction that couples this sector with the valence sector in the light-cone gauge is   written  below explicitly with the momentum arguments: 
\begin{equation}
  \begin{aligned}
    &\Big[V\psi^{(s_q, s_{\bar{q}})}_{q\bar{q};(i_q,i_{\bar{q}})}
   \Big] = 
   \frac{g_s \sqrt{2}}{x_q + x_g} \sum_{i_1 s_1}T^a_{i_q i_1}W^{(s_q,s_1)}_\lambda(p_q, p_g)
    \\
    & \times \psi^{(s_1, s_{\bar{q}})}_{q\bar{q};(i_1, i_{\bar{q}})}(x_{\bar{q}}, {\vec p}_{\bar{q}\perp}) -  \frac{g_s \sqrt{2}}{x_{\bar{q}} + x_g} \\
    & \times \sum_{i_1 s_1}T^a_{i_1 i_{\bar{q}}}\bar{W}^{(s_1, s_{\bar{q}})}_\lambda(p_{\bar{q}}, p_g)\psi^{(s_q, s_1)}_{q\bar{q};(i_q, i_1)}(x_q, \vec{p}_{q\perp})\,,
  \end{aligned} \label{eq:vpsiqqbar}
\end{equation}
 where
$p_{q(\bar q)}\equiv\{ p^+_{q(\bar q)},\vec p_{q(\bar q)\perp}\}$ and $p_g\equiv\{ p^+_g,\vec p_{g\perp}\}$. 
It should be noted that the arguments of the valence wave function correspond to those of the spectator quark/antiquark in the gluon radiation process.

The basic quark-gluon-quark matrix element is given by
\begin{equation}
\label{Eq:W_me}
W^{(s_q,s_1)}_\lambda(p_q, p_g) = \tfrac{1}{\sqrt{2}}\bar{u}_{s_q}(p_q)\,\slashed{\varepsilon}^*_\lambda(p_g)\,u_{s_1}(p_q + p_g)\,,
\end{equation}
and the corresponding matrix element for antiquarks reads
\begin{equation}
\label{Eq:Wb_me}
  \bar{W}^{(s_1, s_{\bar{q}})}_\lambda(p_{\bar{q}}, p_g) = \tfrac{1}{\sqrt{2}}\bar{v}_{s_1}(p_{\bar{q}}+ p_g)\,\slashed{\varepsilon}^*_\lambda(p_g)\,v_{s_{\bar{q}}}(p_{\bar{q}})\,,
\end{equation}
 where $u_{s_q}$ and $v_{s_{\bar q}}$ are the light-cone helicity spinors.

 The matrix elements  \eqref{Eq:W_me} and \eqref{Eq:Wb_me} have been tabulated in Ref.~\cite{Brodsky:1997de}.  For clarity, we write  the spin-flip matrix elements:
\begin{equation}
\begin{aligned}
W^{(+,-)}_\lambda(p_q, p_g)   &= m_f\frac{x_1 - x_q}{\sqrt{x_1 x_q}}\delta_{\lambda, -}\,, \\
 W^{(-,+)}_\lambda(p_q, p_g)   &= -m_f\frac{x_1 - x_q}{\sqrt{x_1 x_q}}\delta_{\lambda, +}\,, \\
\bar{W}^{(+,-)}_\lambda(p_{\bar{q}}, p_g)   &= \lambda m_f\frac{x_1 - x_{\bar{q}}}{\sqrt{x_1 x_{\bar{q}}}}\,, \\
 \bar{W}^{(-,+)}_\lambda(p_{\bar{q}}, p_g)   &= 0\,,
\end{aligned}
\end{equation}
where $x_1 = x_q + x_g \:(x_{\bar{q}} + x_g)$ for (anti)quark matrix elements. The parameter $m_f$ controls their magnitude which, in BLFQ, rules the split between the pion and rho meson masses, lowering the  pion to its small mass in the hadronic scale. 

The  contribution from the $q\bar{q}g$ sector to the quark PDF is
\begin{equation}
\label{Eq:u_q}
\begin{aligned}
 \Delta u_q(x_q) &= \frac{1}{(2\pi)^6} \sum_{i_q s_q}\sum_{i_{\bar{q}} s_{\bar{q}}}\sum_{\lambda a} \int \frac{d^2p_{q\perp}d^2p_{g\perp}dx_g}{4x_qx_g(1 -x_q - x_g )}\\
  & \times|\psi^{(s_q,s_{\bar{q}},\lambda)}_{q\bar{q}g;(i_q,i_{\bar{q}},a)}(\lbrace x, \vec{p}_\perp \rbrace)|^2\, . 
\end{aligned}
\end{equation}
The corresponding expression for the antiquark PDF is obtained through the exchange $q \rightarrow \bar{q}$.
The gluon PDF is similarly given by
\begin{equation}
\label{Eq:u_g}
\begin{aligned}
  u_g(x_g) &= \frac{1}{(2\pi)^6} \sum_{i_q s_q}\sum_{i_{\bar{q}} s_{\bar{q}}}\sum_{\lambda a} \int \frac{d^2p_{q\perp}dx_qd^2p_{g\perp}}{4x_qx_{g}(1 -x_q - x_g)}\\
  & \times|\psi^{(s_q,s_{\bar{q}},\lambda)}_{q\bar{q}g;(i_q,i_{\bar{q}},a)}(\lbrace x, \vec{p}_\perp \rbrace)|^2\,.
\end{aligned}
\end{equation}

The summed squared $q\bar{q}g$ LF wave function entering Eqs.~\eqref{Eq:u_q} and \eqref{Eq:u_g} takes the form  
\begin{equation}
  \begin{aligned}\label{eq:pdfgluon}
   &\sum_{i_q s_q}\sum_{i_{\bar{q}} s_{\bar{q}}}\sum_{\lambda a} |\psi^{(s_q,s_{\bar{q}},\lambda)}_{q\bar{q}g;(i_q,i_{\bar{q}},a)}(\lbrace x, \vec{p}_\perp \rbrace)|^2 \\
   &= \frac{N(N - 1)g_s^2}{(M^2_\pi - M^2_{0,q\bar{q}g})^2} \sum_{s_q s_{\bar{q}}}\sum_{\lambda}\sum_{s_1 s_2} \\ & \Biggl
    \lbrace \frac{[W^{(s_q,s_1)}_\lambda(p_q, p_g)]^*[W^{(s_q,s_2)}_\lambda(p_q, p_g)]}{(x_q + x_g)^2} \\
    & \times[\psi^{(s_1, s_{\bar{q}})}_{q\bar{q}}(x_{\bar{q}}, \vec p_{\bar{q}\perp})]^* \psi^{(s_2, s_{\bar{q}})}_{q\bar{q}}(x_{\bar{q}}, \vec p_{\bar{q}\perp}) \\
    & + \frac{[\bar{W}^{(s_1, s_{\bar{q}})}_\lambda(p_{\bar{q}}, p_g)]^*[\bar{W}^{(s_2,s_{\bar{q})}}_\lambda(p_{\bar{q}}, p_g)]}{(x_{\bar{q}} + x_g)^2} \\
    & \times [\psi^{(s_q,  s_1)}_{q\bar{q}}(x_q, \vec p_{q\perp})]^* \psi^{(s_q, s_2)}_{q\bar{q}}(x_q, \vec p_{q\perp}) \\
    &  - 2Re\Bigl[\frac{[W^{(s_q,s_1)}_\lambda(p_q, p_g)]^*\bar{W}^{(s_2,s_{\bar{q})}}_\lambda(p_{\bar{q}}, p_g)}{(x_q + x_g)(x_{\bar{q}}+ x_g)}  \\ & \times [\psi^{(s_1, s_{\bar{q}})}_{q\bar{q}}(x_{\bar{q}}, \vec p_{\bar{q}\perp})]^* \psi^{(s_q, s_2)}_{q\bar{q}}(x_q, \vec p_{q\perp})    \Bigr] \Biggr\rbrace\,,
    \end{aligned}
\end{equation}
 where we have chosen to denote in the valence wave function the spectator quark or antiquark momenta in the gluon radiation process.
In the present work it will be assumed that the valence wave function is dominated by spin-antialigned component, $\psi^{(+, -)}_{q\bar{q}}=-\psi^{(-, +)}_{q\bar{q}}$~(see e.g.~\cite{dePaula:2020qna}), and the aligned contribution will thus be neglected. 


\begin{table}[t!]
    \caption{Quark mass in kinetic part of Hamiltonian, quark mass in one-gluon-exchange interaction and gluon mass for the three adopted parameter sets in GeV. The rightmost column displays the probability of the $q\bar{q}g$ component.}
    \label{Tab:params}
\label{Tab:param_sets}
\vspace{0.2cm}
    \centering
    \begin{tabular}{c c c c c}
    \toprule
      Model  & $m_q$[GeV] & $m_f$[GeV] & $m_g$[GeV] & $1$-$P_{val}$  \\
       \midrule
       I  & 0.390 & 0.390 & 0.600 & 0.508 \\
       II & 0.390 & 5.69 & 0.600 & 0.508 \\
       III & 0.255 & 0.255 & 0.638 & 0.300 \\
        \bottomrule       
    \end{tabular}
\end{table}

\section{Results and discussion}
In the present work, the gluon contributions to the pion PDFs are studied. In particular, we compute the $q\bar{q}g$ contribution to the quark PDF and gluon PDF by using the formalism outlined in Sec.~\ref{Sec:PLFH}. The results with an input model valence wave function will then be compared to those of BLFQ \cite{Lan:2021wok} and those of BSE.

The varied inputs of our model are the quark mass $m_q$ entering the kinetic part (see Eq.~\eqref{Eq:M0_qqg}), the quark mass entering the interaction ($m_f$) and the gluon mass $m_g$.  In the present study we consider three different parameter sets which are given in Table \ref{Tab:params}. Namely, in Model I and II we use $m_q$ and $m_g$ from BLFQ \cite{Lan:2021wok}. However, in the first case $m_f=m_q$ instead of $m_f = 5.69$ GeV that gives experimental mass splitting between $\pi$ and $\rho$. The parameter $m_f$ is an \textit{effective} quark mass, incorporating non-perturbative effects like gluonic contributions within the light-front QCD Hamiltonian. While it appears large compared to the physical quark mass, it captures the dressed quark's collective behavior, ensuring consistency with observed hadronic spectra. The last set is using the values of the masses as in the BSE \cite{dePaula:2020qna}. The coupling constant $g_s=1.92$ and other parameters of BLFQ are taken from Ref.~\cite{Lan:2021wok}.

\begin{figure}[thp!]
    \centering
    \includegraphics[scale=0.34]{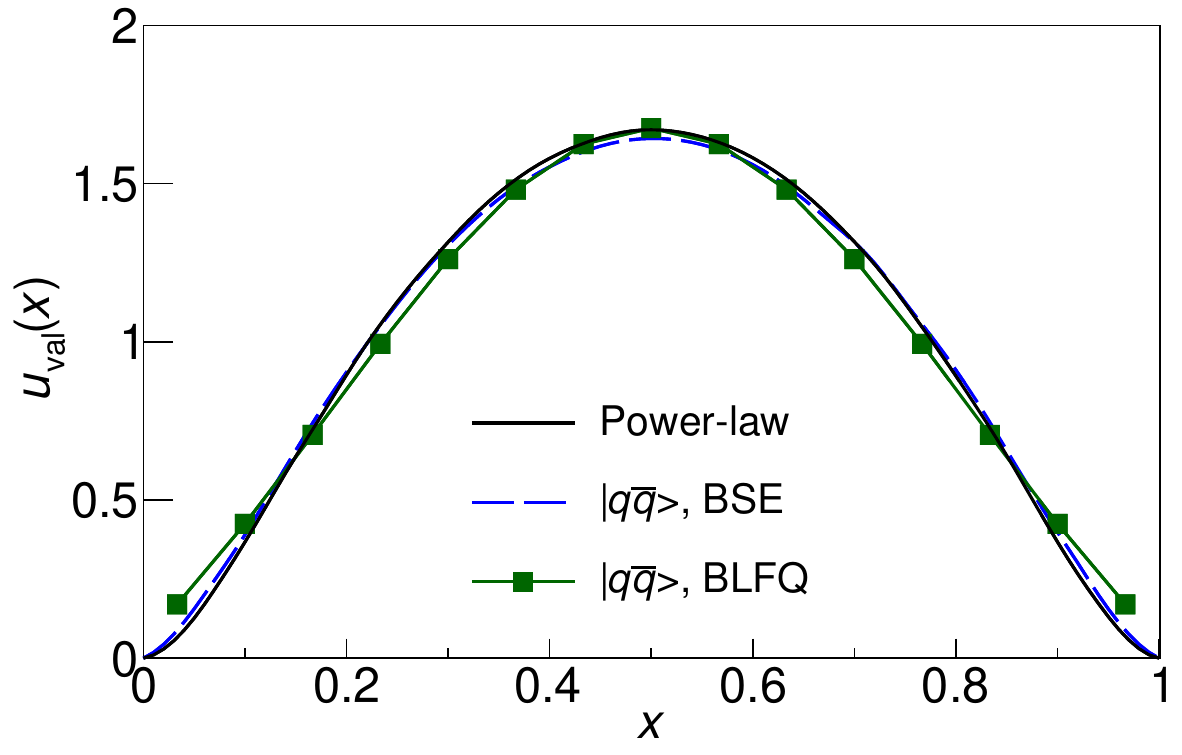}
    \includegraphics[scale=0.34]{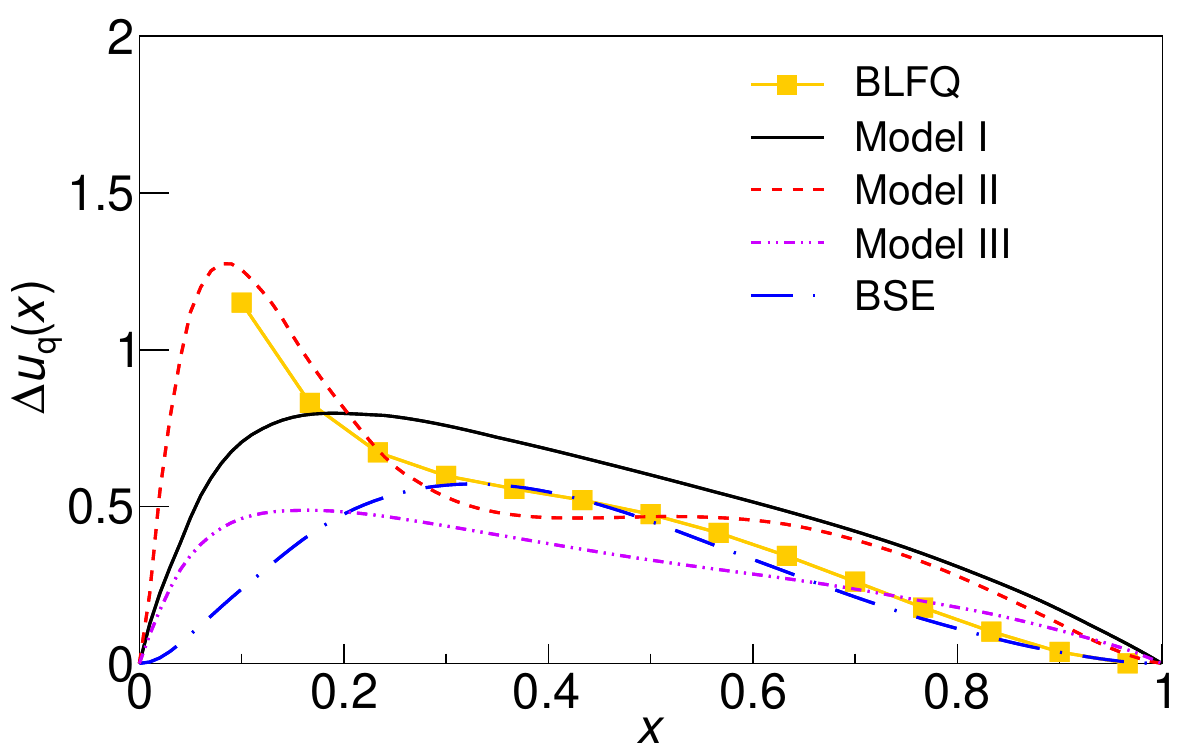}   
    \includegraphics[scale=0.34]{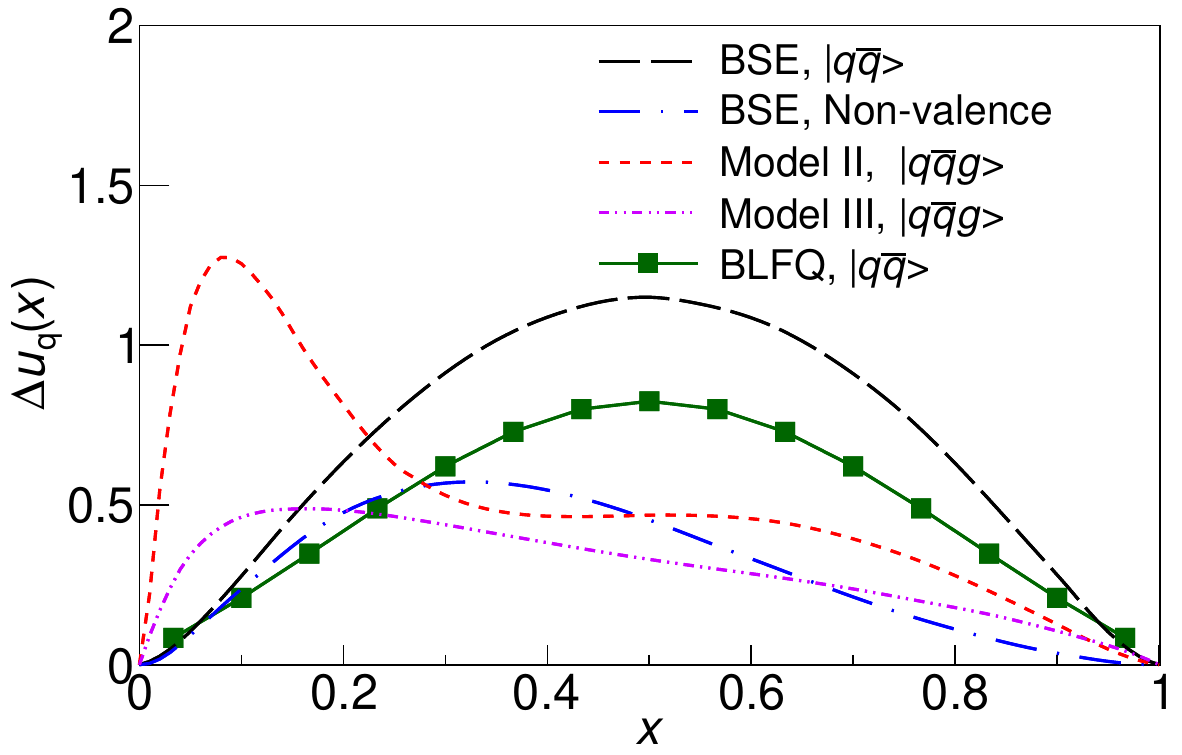} 
    \caption{Upper panel: The pion's valence quark PDF calculated using the power-law form of the pion wave function, Eq.~\eqref{power-law}, compared with the results of BSE (long dashed line) and the BLFQ (squares connected with solid line). The valence quark PDFs are rescaled to unit norm to facilitate a qualitative comparison between the BLFQ and BSE approaches. Middle panel: Contribution from the $q\bar{q}g$ sector to the quark PDF calculated within the perturbative approach with the parameter sets I (solid line), II (dashed line) and III (dash-double dotted line) compared with the results of BSE (long dashed line) and the BLFQ (squares connected with solid line).
    Lower panel: The valence and $q\bar{q}$ contribution to the pion's quark PDF calculated within the BLFQ (squares connected with solid line) and BSE (long dashed line) frameworks, respectively are compared with the $q\bar{q}g$ for Model II (dash line) and III (dash-double dotted line). The long dashed-single dotted line represents the contribution from the non-valence sectors to the quark PDF computed within BSE.}
    \label{fig:u_q}
\end{figure}

For simplicity  in the present study we will use a power-law form~\cite{Schlumpf_94}:
\begin{equation}\label{power-law}
  \psi_{\rm pl}(x, \vec p_{\perp}) = N[1 +(A_{0,{\rm eff}}(x, \vec p_{\perp})/4 - m^2_q)/\beta^2]^{-s},
\end{equation}
in the place of the valence amplitude. 
$N$ is a normalization constant and the parameters $s$ and $\beta$ will be determined through a fit to either BLFQ or BSE results. Moreover, the effective function $A_{0,{\rm eff}}(x, \vec p_{\perp})$ 
is chosen as
\begin{equation}\label{eq:A0}
  A_{0,\text{eff}}(x_q,\vec p_{q\perp}) = \frac{p^2_{q\perp} + m^2_q}{x_q} + \frac{p_{\bar q\perp}^2 + m^2_q}{ x_{\bar q}}\, .
\end{equation}
In the actual calculations of the $q\bar q g$ contribution to the momentum distributions we have used: $$\vec p_{\bar q}=-\vec p_{q\perp}-\vec p_{g\perp}\quad \text{and} \quad x_{\bar q}=1-x_q-x_g\,,$$ corresponding to the final momentum of the antiquark after the gluon is radiated and $q$ is the spectator quark.  When $\bar q$ is the spectator an analogous expression is used, by exchanging the momenta of the quark with the antiquark. The power-law form in Eq.~\eqref{power-law} is chosen for its simplicity, physical relevance, and consistency with perturbative QCD predictions. It ensures smooth behavior at small transverse momenta and a power-law falloff at large transverse momenta, with the exponent $s$ often determined phenomenologically or from constraints like quark counting rules. This form, widely used in studies of hadron structure, distribution amplitudes, and form factors, balances simplicity and physical relevance, making it a natural choice for our analysis~\cite{Schlumpf_94,Hwang:2010hw,Hwang:2001wu,Geng:2016pyr}.
Although Eq.~\eqref{eq:A0} does not correspond to the standard formulation used in valence wave function models, where the free mass operator of the $q\bar{q}$ system is typically employed, the amplitude expressed in Eq.~\eqref{power-law}, when substituted into Eq.~\eqref{eq:pdfgluon}, reasonably accounts for the contribution of the $q\bar{q}g$ sector to the quark parton distribution obtained with BLFQ, as we will demonstrate.
This simple recipe takes into account the damping of the loop integral in Eq.~\eqref{Eq:u_q} close to $x_g\to 1$, and the decrease of the gluon distribution in Eq.~\eqref{Eq:u_g} close to the end-point. We plan in the future to use a more general form of the valence wave function which reflects the dynamical content of the BLFQ Hamiltonian and BS equation.

Note that we fit $s$ and $\beta$ parameters with the contribution of the valence state to the quark PDF.  In this case, the function $ A_{0,\text{eff}}(x_q,\vec p_{q\perp})$ reduces to the standard mass squared function, namely:
\begin{equation}\label{eq:m02qqbar}
M^2_{0,q\bar q}(x, \vec p_{\perp}) = 
\frac{\vec p^{\,2}_{\perp} + m^2_q}{x(1-x)}\,,
\end{equation}
which was used in the fitting of the $q\bar q$ leading Fock-sector  momentum distributions from the valence state obtained with BLFQ and BSE calculations.

We find that the parameters $s=1.4$ and $\beta/m_q=1.16$  reproduce well the valence PDF of both the BLFQ and the BSE as shown in the upper panel of Fig.~\ref{fig:u_q}.

\begin{figure}[t]
    \centering
    \includegraphics[scale=0.34]{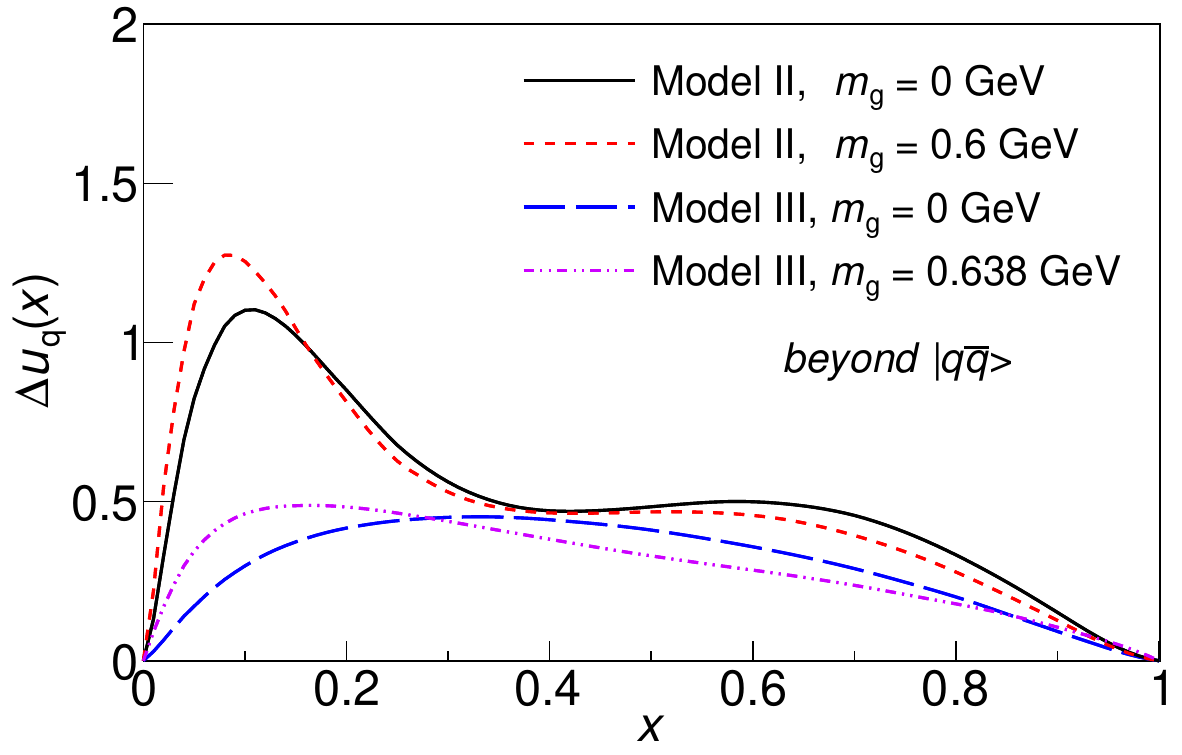}
    \includegraphics[scale=0.34]{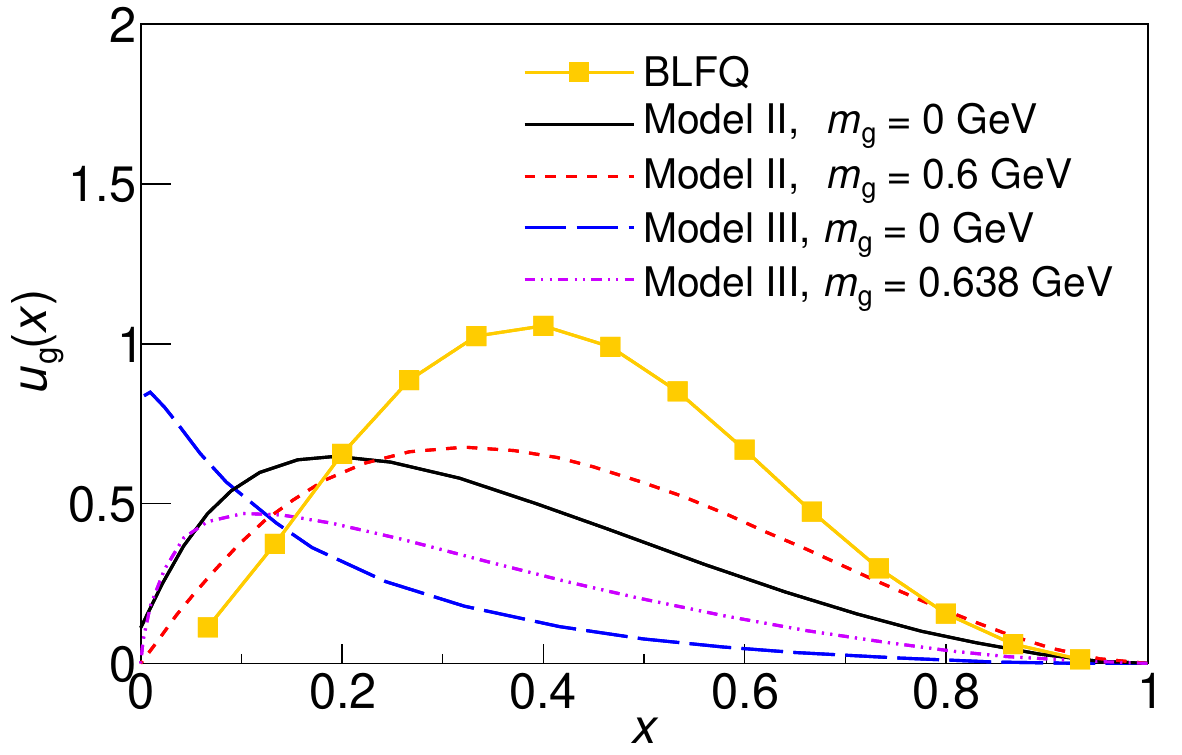}  
    \caption{Upper panel: Contribution to quark PDF from $q\bar{q}g$ sector for Model II but with $m_g = 0$ GeV (solid line), Model II with $m_g$ of Table \ref{Tab:params} (dashed line), Model III with $m_g=0$ GeV (long dashed line) and Model III with $m_g$ from Table \ref{Tab:params} (long dash-double dotted line). Lower panel: Gluon PDF for Model II but with $m_g = 0$ GeV (solid line), Model II with $m_g$ of Table \ref{Tab:params} (dashed line), Model III with $m_g=0$ GeV (long dashed line) and Model III with $m_g$ from Table \ref{Tab:params} (long dash-double dotted line). For comparison, the gluon PDF calculated with BLFQ (squares connected with solid line)  is also shown.}
    \label{fig:u_q_u_g_m_g}
\end{figure}

In the middle panel of Fig.~\ref{fig:u_q}, we compare the results for the $q\bar{q}g$ contribution to the quark PDF for the sets I, II, III with the BSE calculation for the beyond-valence contribution and also the result of the BLFQ. From the figure, it is seen that the Model II  qualitatively agrees with the BLFQ, as it should. Namely, the large bump at low-$x$ is reproduced. By comparing the results for Model I and II, it can be concluded that the mentioned bump is related to the large value of $m_f = 5.69$ GeV, used in Model II. 
Note that the reproduction of the spectrum, i.e.~a small pion mass, requires the large value of $m_f$~\cite{Lan:2021wok}. Furthermore, it can also be  seen in the middle panel of Fig.~\ref{fig:u_q} that the BSE result differs quite significantly from the Model III. But, one should notice in such a comparison that the BSE result contains an infinite number of contributions of the form $q\bar{q}ng$ where the number $n=1, \cdots, \infty$ is the number of gluons. Additionally, the BSE calculation was performed in the Feynman gauge. The discrepancy between Model II and the BLFQ result can presumably be explained by the use of a simple analytical form in the numerical calculations and the fact that the BLFQ is using a discretization of the longitudinal fractions not used in the perturbative method developed in this work. The significance of the second Fock sector at small-$x$ is evident within our BLFQ framework, especially given the large $m_f$ value in Model II. The discrepancies with the BSE results underscore the strong dependence of the low-$x$ behavior on the specific treatment of gluonic contributions, the choice of gauge, and possibly, the adoption of the large-$N_c$ limit of QCD. 
This motivates to apply the BSE in the LC gauge ~\cite{Liu:1992dg}  for future studies of the pion.

The valence and $q\bar{q}$ contribution to the PDF computed within the BLFQ and BSE frameworks, respectively is compared with the $q\bar{q}g$ for Model II and III in the lower panel of Fig.~\ref{fig:u_q}. It can be concluded that the second Fock sector is  important at small-$x$. As expected, the valence component dominates at larger values of $x$. 

We now examine the contributions of the higher Fock component to the first moments of the quark, antiquark, and gluon PDFs obtained from Model III, specifically $\langle x^{q\bar qg}_q \rangle = \langle x^{q\bar qg}_{\bar q} \rangle$ and $\langle x^{q\bar qg}_g \rangle$, which attain values close to $0.1$. These results are in reasonable agreement with the values computed using the BSE pion model, which are $\langle x^{q\bar qg}_q \rangle = \langle x^{q\bar qg}_{\bar q} \rangle = 0.12$ and $\langle x^{q\bar qg}_g \rangle = 0.06~$\cite{dePaula:2023ver}. This suggests that the difference in gauge choice does not significantly affect the qualitative results, allowing for a comparison of the BSE results with the BLFQ results in the light-cone gauge. 

We now examine the impact of the gluon mass $m_g$ on the quark and gluon PDFs in the pion. The results for the $q\bar{q}g$ contribution to the quark PDF are shown in the upper panel of Fig.~\ref{fig:u_q_u_g_m_g} for Model II and III that use two different values of $m_g$, i.e.~the values given in Table~\ref{Tab:params}, as well as for a vanishing gluon mass. %
In addition, as seen in the figure, an increase of the gluon mass leads to a shift of the quark PDF to lower values of $x$ for both models. However, for Model III with $m_f = m_q$ the effect is more pronounced compared to Model II having a large value of $m_f$. Similarly, we show in the lower panel of Fig.~\ref{fig:u_q_u_g_m_g} the results for the gluon PDF. The behavior is now the opposite, i.e.,~a larger $m_g$ gives a gluon PDF shifted towards larger-$x$. In this figure,  we also compare those results with the gluon PDF computed within BLFQ. 
The perturbative results agree qualitatively with those from BLFQ, although the latter framework provides a PDF that is slightly shifted towards higher values of $x$.
Model II has a distribution peaked much more to the right compared to Model III, i.e.~increasing the mass $m_f$ leads to a larger $\langle x \rangle_g$ of the gluon. 

\begin{figure}[t]
    \centering
    \includegraphics[scale=0.34]{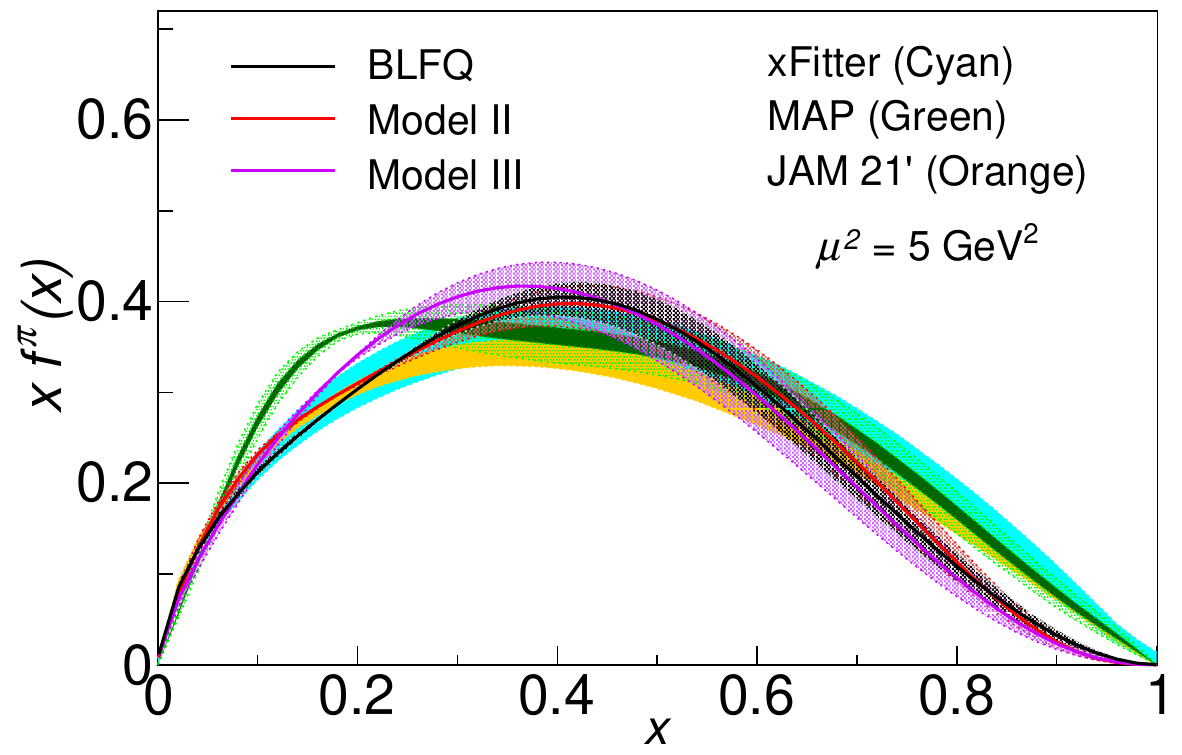}
    \includegraphics[scale=0.34]{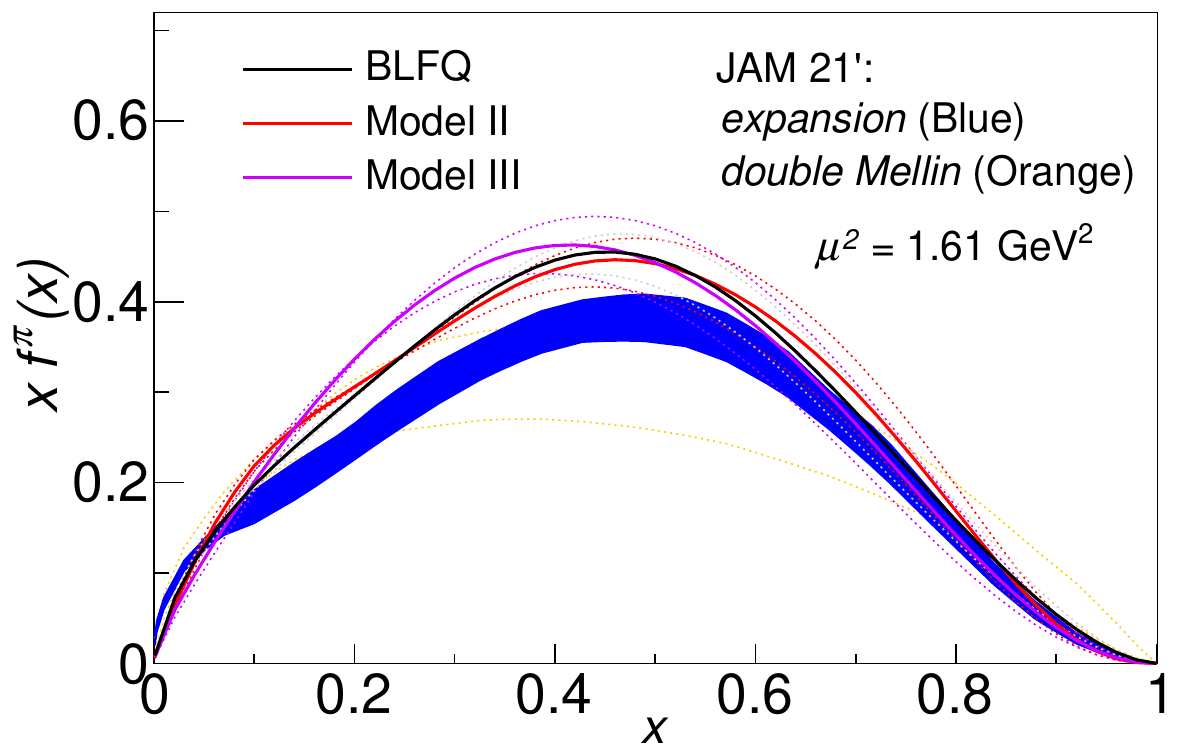}
    \includegraphics[scale=0.34]{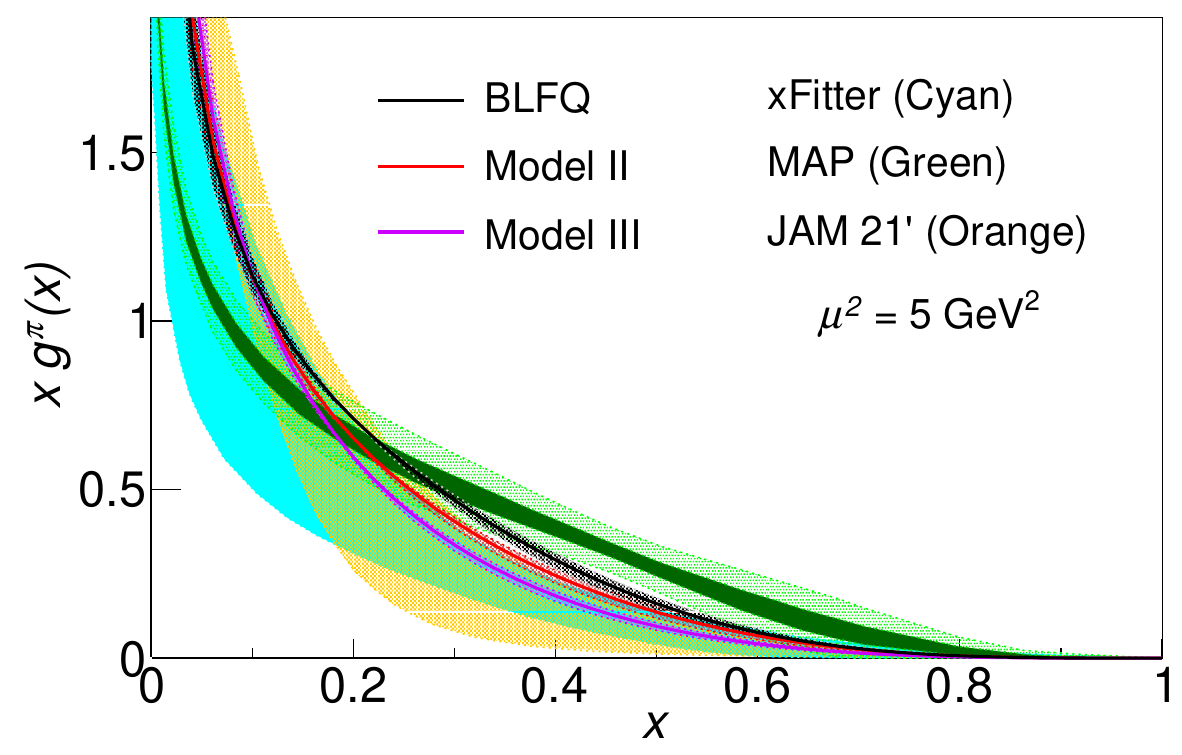}
    \caption{The PDFs of the pion. Upper panel: The black (original BLFQ),  red (Model II), and violet (Model III) lines are our valence quark PDFs evolved from the initial scales using the NNLO DGLAP equations to  $5~\mathrm{GeV}^2$. Our results are compared with the global QCD analyses by JAM21~\cite{Barry:2021osv} (orange band), MAP~\cite{Pasquini:2023aaf} (green band), and xFitter~\cite{Novikov:2020snp} (cyan band) Collaborations. Middle panel: We compare our evolved valence quark PDFs with two different analyses from JAM21 Collaboration at  $1.61~\mathrm{GeV}^2$. The blue and orange bands correspond to the JAM21 analyses, which use the expansion method and the double Mellin method, respectively for threshold resummation on Drell-Yan cross sections at next-to-leading log accuracy~\cite{Barry:2021osv}. Lower panel: Our results for the pion gluon PDF at $5$ GeV$^2$ are compared with those global QCD analyses as mentioned in the top panel.}
    \label{fig_pdf}
\end{figure}

The upper panel of Fig.~\ref{fig_pdf} presents our pion PDF results, comparing the evolved valence quark distribution with the global QCD analyses by JAM21~\cite{Barry:2021osv}, MAP~\cite{Pasquini:2023aaf}, and xFitter~\cite{Novikov:2020snp} Collaborations. We determine the initial scale $\mu_0^2$ by requiring that the evolved results reproduce the total first moments of the valence quark and antiquark distributions from the global QCD analysis, $\langle x \rangle_{\text{valence}} = 0.48 \pm 0.01$ at $\mu^2 = 5~\text{GeV}^2$~\cite{Barry:2018ort}. This yields the initial scales: $\mu_{\text{BLFQ}0}^2 = 0.34 \pm 0.03~\text{GeV}^2$ for BLFQ~\cite{Lan:2021wok}, $\mu_{\text{II}0}^2 = 0.30 \pm 0.03~\text{GeV}^2$ for Model II, and $\mu_{\text{III}0}^2 = 0.26 \pm 0.03~\text{GeV}^2$ for Model III. Our evolved PDFs include a $10\%$ uncertainty band at the initial scale.  We observe that our valence quark PDFs show overall good consistency with these global QCD analyses. 

Our predictions agree well with the JAM21 analysis when it employs the expansion method for threshold resummation on Drell-Yan cross sections at next-to-leading log accuracy~\cite{Barry:2021osv}, as shown in the middle panel of Fig.~\ref{fig_pdf}. However, at large-$x$, our results deviate from the JAM21 analysis when it uses the double Mellin method for resummation effects~\cite{Barry:2021osv}. Our pion valence PDF falls as $(1-x)^{1.77}$, slightly slower than perturbative QCD's $(1-x)^2$~\cite{Berger:1979du} and DSE results~\cite{Hecht:2000xa}, consistent with Refs.~\cite{Chen:2016sno,Aicher:2010cb,Bednar:2018mtf} and the JAM21 analysis with the expansion method~\cite{Barry:2021osv}. In contrast, the JAM21 global analysis with the double Mellin method shows a linear falloff, $(1-x)^{1}$~\cite{Barry:2021osv}. Meanwhile, we find good consistency between our results for the gluon PDF and the global QCD analyses~\cite{Pasquini:2023aaf}, as shown in the lower panel of Fig.~\ref{fig_pdf}.

As a final remark, we verified that the BLFQ results for the contribution of the higher Fock component to the quark and gluon PDFs, computed using the Model I and III parameters from Table~\ref{Tab:params}, are consistent with the results shown in Figs.~\ref{fig:u_q} and \ref{fig:u_q_u_g_m_g}, after renormalizing to $0.508$ and $0.3$, respectively. These results were obtained with the present framework by applying Eq.~\eqref{Eq:psi_qqg} to compute the quark-antiquark-gluon component of the pion state from the valence wave function. Note that the values $0.508$ and $0.3$ represent the probabilities of the $q\bar{q}g$ Fock component in the BLFQ framework~\cite{Lan:2021wok,Zhu:2024awq} and the non-valence components in the BSE framework~\cite{dePaula:2022pcb}, respectively, computed from their respective LFWFs, as indicated in Table \ref{Tab:param_sets}.

\section{Conclusion}
In this study, we analyzed the pion $|q\bar{q}g\rangle$ contributions to the quark PDF beyond the leading $|q\bar{q}\rangle$ Fock sector within the BLFQ framework and developed a perturbative method to generate the $|q\bar{q}g\rangle$ contributions from the valence Fock component. We compared these results with direct BLFQ and Minkowski-space BSE calculations, exploring variations in parameters such as the gluon mass and constituent quark masses.
 In the BLFQ case, we identified the  effect of the dynamical chiral symmetry breaking in the pion quark PDF, namely the  enhancement of the spin-flip matrix element impacts the $|q\bar qg\rangle$ Fock-component of the LF wave function  and  associated gluon PDF. In particular, we explicitly showed that  the low-$x$ peaked contribution to the quark PDF is directly associated with the large spin-flip matrix element that drives the $\pi-\rho$ mass splitting.  
The proposed framework was also used to investigate the effect of the dressed gluon mass for a given valence state at the pion scale. It is shown that the gluon PDF experiences a significant change when the gluon mass vanishes.

The explored framework in the light-cone gauge can be applied to other pion models of the valence state to build the $|q\bar qg\rangle$ component and eventually provide insights into the roles of the higher Fock-components. 
As it stands, the present approach provides a practical method for computing the quark and gluon parton distributions, taking into account the higher Fock component of the pion state. We have compared the calculations of the pion quark and gluon PDFs, including the contribution from the $q\bar{q}g$ component computed using the proposed operator acting on the valence wave function, with the BLFQ results, as well as with those obtained from the solution of the BSE in Minkowski space.

We can foresee some further steps to apply our methodologies. For example,  one may look for the $q\bar q g$ component extracted from phenomenological parametrizations, like the one developed in Ref.~\cite{Pasquini:2023aaf}, to further support the enhancement of the spin-flip matrix element. Meanwhile, this method could be applied to separate the $q\bar q g$ component starting with the valence pion wave function obtained within DS/BSE approaches, although in covariant gauges. These are future challenges in the perspective to deepen our understanding of the pion LF wave function in the Fock-space with dressed constituents. 

The other direct manifestation of the higher Fock component with massive gluons appears in the clear separation between the peaks of the gluon PDF and the contribution to the quark PDF when the spin-flip matrix element is enhanced to provide the $\pi-\rho$ splitting. On the other side, with parameters from the pion BS model in Minkowski space, where the enhancement of the spin-flip matrix element is quite mild, the gluon and quark PDF from the $q\bar q g$ component peaks around the same position at $x\sim 0.15$. The  Fock components beyond the valence from  the BS model in Minkowski space provide a contribution to the quark PDF that is peaked around $x\sim 0.3$. The source of this difference could be associated with the extension of the quark-gluon vertex, which was tested here in a qualitative way, providing the shift  from $x\sim 0.15$ to around $\sim 0.3$ of the peak in the PDF for the contribution  of the $q\bar q g$ state. The application to the nucleon to compute the $qqqg$ component is another challenge that could begin with the recent BLFQ  results  for the proton~\cite{Xu:2021wwj,Xu:2023nqv}, and using proton valence models from Minkowski space dynamics (see e.g. ~\cite{Ydrefors:2022bhq}).

Based on these inter-comparisons, we speculate that increasing the number of gluons in the retained Fock sectors within the BLFQ approach will: (i) drive the $\pi-\rho$ splitting, (ii) reduce the effective vertex mass of fermions, and (iii) increase the non-valence sector contributions to the quark PDF in the low-$x$ region.

\begin{acknowledgments}
The authors would like to thank Dr.~Emanuel Ydrefors for his assistance in resolving issues encountered at all stages of this work. J.L. is supported by Special Research Assistant Funding Project, Chinese Academy of Sciences, by the Natural Science Foundation of Gansu Province, China, Grant No.23JRRA631, and by National Natural Science Foundation of China, Grant No. 12305095. C.M. is supported by new faculty start up funding the Institute of Modern Physics, Chinese Academy of Sciences, Grants No. E129952YR0. C.M. also thanks the Chinese Academy of Sciences Presidents International Fellowship Initiative for the support via Grants No. 2021PM0023. 
X.Z. is supported by new faculty startup funding by the Institute of Modern Physics, Chinese Academy of Sciences, by Key Research Program of Frontier Sciences, Chinese Academy of Sciences, Grant No. ZDB-SLY-7020, by the Natural Science Foundation of Gansu Province, China, Grant No. 20JR10RA067, by the Foundation for Key Talents of Gansu Province, by the Central Funds Guiding the Local Science and Technology Development of Gansu Province, Grant No. 22ZY1QA006, by National Natural Science Foundation of China, Grant No. 12375143, by National
Key R\&D Program of China, Grant No. 2023YFA1606903 and by the Strategic Priority Research Program of the Chinese Academy of Sciences, Grant No. XDB34000000. This research is supported by Gansu International Collaboration and Talents Recruitment Base of Particle Physics (2023-2027), by the Senior Scientist Program funded by Gansu Province, Grant No. 25RCKA008, and supported by the International Partnership Program of Chinese Academy of Sciences, Grant No.016GJHZ2022103FN.
This work is a part of the project INCT-FNA \#464898/2014-5.
This study was financed in part by Conselho  Nacional de Desenvolvimento Cient\'{i}fico e  Tecnol\'{o}gico (CNPq) under the grant 306834/2022-7 (TF).  We  thank the FAPESP Thematic  grant     \#2019/07767-1.  
J. P. V. is supported by the U.S. Department of Energy under Grant No. DE-SC0023692. A portion of the computational resources were also provided by Taiyuan Advanced Computing Center.
\end{acknowledgments}

\bibliography{references.bib}
\end{document}